\documentclass[review,preprint,authoryear]{elsarticle}

\usepackage{graphicx}
\usepackage{amssymb}
\usepackage{multirow}
\usepackage{mathptmx}
\usepackage{times}
\usepackage{epsfig}
\usepackage{graphicx}
\usepackage{amsmath}
\usepackage{amssymb}
\usepackage[table,xcdraw]{xcolor}
\usepackage{caption}
\usepackage{todonotes}
\usepackage{xcolor}
\usepackage{url}
\usepackage{subfig}
\usepackage{blindtext}
\usepackage{latexsym}
\usepackage{lineno,hyperref}
\usepackage{framed,multirow}
\usepackage{booktabs}

\usepackage{amssymb}
\usepackage{latexsym}

\usepackage{hyperref}

\journal{Journal of \LaTeX\ Templates}

\definecolor{newcolor}{rgb}{.8,.349,.1}

\journal{Expert Systems with Applications}

\begin{document}

\begin{frontmatter}

\title{SLSNet: Skin lesion segmentation using a lightweight generative adversarial network}


\author[add1,add2]{Md. Mostafa Kamal Sarker\corref{mycorrespondingauthor}}
\ead{mdmostafakamalsarker@ub.edu}
\author[add2]{Hatem A.Rashwan}
\ead{hatem.abdellatif@urv.cat}
\author[add3,add9]{Farhan Akram}
\ead{farhan@vim.cau.ac.kr}
\author[add2]{Vivek Kumar Singh}
\ead{vivekkumar.singh@urv.cat}
\author[add2]{Syeda Furruka Banu}
\ead{syeda.samia@hotmail.com}
\author[add6]{ Forhad U. H. Chowdhury}
\ead{drmarufsomc@gmail.com}
\author[add7]{ Kabir Ahmed Choudhury}
\ead{K.Choudhury@warwick.ac.uk }
\author[add4]{ Sylvie Chambon}
\ead{sylvie.chambon@toulouse-inp.fr}
\author[add1,add8]{ Petia Radeva}
\ead{petia.ivanova@ub.edu}
\author[add2]{Domenec Puig}
\ead{domenec.puig@urv.cat}
\author[add2,add5]{Mohamed~Abdel-Nasser}
\ead{mohamed.abdelnasser@urv.cat}

\cortext[mycorrespondingauthor]{Corresponding author}

\address[add1]{Departament de Matemàtiques i Informàtica, University of Barcelona, 08007 Barcelona, Spain.}
\address[add2]{Departament d'Enginyeria Informàtica i Matemàtiques, Universitat Rovira i Virgili, 43007 Tarragona, Spain.}
\address[add3]{Department of Electrical and Computer Engineering, Khalifa University of Science and Technology, 127788 Dubai, UAE.}
\address[add4]{ Institut de Recherche en Informatique de Toulouse, University of Toulouse, 31000 Toulouse, France}
\address[add5]{Department of Electrical Engineering, Aswan University, 81528 Aswan, Egypt.}
\address[add6]{Dhaka Medical College Hospital, 1209 Dhaka, Bangladesh.}
\address[add7]{Warwick Medical School, University of Warwick, Coventry CV4 7HL, UK.}
\address[add8]{Computer Vision Center, University of Barcelona, 08193, Barcelona, Spain.}
\address[add9]{Mil-kin Inc.,  2F, 2-6-1, Otemachi, Chiyoda-ku, Tokyo, Japan.}

\begin{abstract}

The determination of precise skin lesion boundaries in dermoscopic images using automated methods faces many challenges, most importantly, the presence of hair, inconspicuous lesion edges and low contrast in dermoscopic images, and variability in the color, texture and shapes of skin lesions. Existing deep learning-based skin lesion segmentation algorithms are expensive in terms of computational time and memory. Consequently, running such segmentation algorithms requires a powerful GPU and high bandwidth memory, which are not available in dermoscopy devices. Thus, this article aims to achieve precise skin lesion segmentation with minimum resources: a lightweight, efficient generative adversarial network (GAN) model called SLSNet, which combines 1-D kernel factorized networks, position and channel attention, and multiscale aggregation mechanisms with a GAN model. The 1-D kernel factorized network reduces the computational cost of 2D filtering. The position and channel attention modules enhance the discriminative ability between the lesion and non-lesion feature representations in spatial and channel dimensions, respectively. A multiscale block is also used to aggregate the coarse-to-fine features of input skin images and reduce the effect of the artifacts. SLSNet is evaluated on two publicly available datasets: ISBI 2017 and the ISIC 2018. Although SLSNet has only 2.35 million parameters, the experimental results demonstrate that it achieves segmentation results on a par with the state-of-the-art skin lesion segmentation methods with an accuracy of 97.61\%, and Dice and Jaccard similarity coefficients of 90.63\% and 81.98\%, respectively. SLSNet can run at more than 110 frames per second (FPS) in a single GTX1080Ti GPU, which is faster than well-known deep learning-based image segmentation models, such as FCN. Therefore, SLSNet can be used for practical dermoscopic applications.

\end{abstract}


\end{frontmatter}


\section{Introduction}
There are 1.04 million melanoma cases in 2018 in the world according to the World Health Organization (WHO)~\footnote{https://www.who.int/news-room/fact-sheets/detail/cancer}. Over the last decades, the number of patients affected by melanoma or non-melanoma skin cancers has been rapidly increased~\citep{apalla2017skin}. 
With the growth of Artificial Intelligence techniques, computer  vision and image analysis techniques based on computerized non-invasive dermatology are essential for dermatologists for early detection of malignant melanoma~\citep{esteva2017dermatologist} to increase the survival rate and reduce the diagnosis/treatment cost. Therefore, a computer-aided diagnosis (CADx) system is essential to support the dermatologists to explore the images captured by digital dermatoscopes. The main challenges that face skin lesion segmentation methods are: 1) the vast diversity in color, shape, texture, size, irregular, and fuzzy boundaries of lesions, 2) the presence of blood vessels and hairs, and 3) the low contrast between skin tissues~\citep{al2018skin}.In order to cope with these challenges, several approaches have been presented by using traditional image processing algorithms, such as histogram thresholding, unsupervised clustering, and supervised segmentation methods. For more information, the work in~\citep{celebi2015state} has presented  a comprehensive survey for traditional segmentation techniques. However, these approaches yield inaccurate segmentation results when the skin lesions have fuzzy boundaries~\citep{celebi2015state}. Besides, the performance of these methods highly depends on applying different transformations algorithms for improving the inspected images, such as hair removal and contrast enhancement. With the tremendous progress in the field of machine learning, mainly in deep learning techniques, many skin lesion segmentation approaches have been introduced that increased the accuracy of skin lesion segmentation. For instance, the SLSDeep model was proposed in~\citep{sarker2018slsdeep} to segment the skin lesion by using feature pyramid pooling. In~\citep{al2018skin}, a full resolution convolutional networks (FrCN) has been introduced to directly learn the full resolution visual content of the input images without the need to image processing assisted tools to refine them. Additionally, a generative adversarial network (GAN) with an improved loss function, called SegAN, has also been introduced for learning semantic features of skin lesions in multiscale image representations~\citep{xue2018adversarial}. Besides, some lightweight image segmentation models have been used for skin lesion segmentation, such as a lightweight model so-called ENet \citep{paszke2016enet}. However, it has yielded lower accuracy than state-of-the-art lesion segmentation methods. 

Although, some of the aforementioned methods have provided acceptable skin segmentation precision, they have hundreds of million parameters that make them unsuitable for practical applications and cannot be easily transferred to clinical settings, especially with dermatoscopy devices with limited computational and memory resources. Existing lightweight segmentation models like ENet give results lower than state-of-the-art when applied to skin lesion segmentation. Therefore, there is a need for a light-weight skin lesion segmentation model that can yield competitive results to the methods in the literature. In this work, we propose a lightweight GAN model, named MobileGAN, for segmenting melanoma in dermoscopic images. In MobileGAN, we extract low-level skin lesion-relevant features with multiscale convolutional networks. MobileGAN also adopts a 1-D kernel factorized network to minimize the computational cost and the resources. Moreover, we exploit the position and channel attention mechanisms to promote skin lesion feature representations. Consequently, we can summarize the main contributions of this work as follows: 

\begin{itemize}
\item [$\bullet$]  The main objective is to develop an efficient skin lesion segmentation at low computational cost with keeping high precise segmentation competitive to state-of-the-art models. Hence, we propose a lightweight and fully automatic skin lesion segmentation model, called MobileGAN. 
\item [$\bullet$]  
A multiscale aggregation mechanism is introduced in MobileGAN to extract skin lesions relevant features at different scale representations and cope with the variability of lesion shapes. In order to minimize the number of the trained parameters, the 1-D kernel factorized networks~\citep{romera2018erfnet} are exploited instead of the traditional 2D convolution networks.
\item [$\bullet$] 
We adopt the position and channel attention mechanisms ~\citep{fu2018dual} to capture the correlation between the channel and spatial features responses of the proposed network and enhance the discriminant ability in between lesion and non-lesion feature representations.

\item [$\bullet$]  We propose the use of binary cross-entropy, Jaccard index, and $L_1$-norm to formulate a loss function to address the challenges accompanied by artifacts existing in dermoscopic images.
\end{itemize} 

The organization of this paper is as follows. Section II discusses recent skin lesion segmentation methods based on classical computer vision and deep learning techniques. The architecture of the proposed model and the experimental results are explained in Section III and IV respectively. In turn, Section V presents a general discussion about the study and its results. Finally, Section VI concludes and suggests some ongoing and future lines of this research.

\section{Related work}
Computer vision and machine learning researchers have presented many approaches for skin lesion segmentation based on classical and deep learning techniques. Below, we present and discuss the most common skin lesion segmentation methods.

\textbf{Classical computer vision-based approaches:} The traditional segmentation approaches are mainly include thresholding, active contour~\citep{silveira2009comparison}, region growing~\citep{rahman2016developing}, and unsupervised learning, e.g., clustering~\citep{agarwal2017automated} for melanoma segmentation. Adaptive thresholding and region growing based methods are proposed in~\citep{rahman2016developing} for skin lesion segmentation. A machine learning technqiues, such as Support Vector Machine (SVM), are fed by the segmented regions to classify the type of skin lesions. The thresholding-based approaches only yield good results with apparent boundaries, which is not a common scenario in this task. Regarding to contour-based methods, such as adaptive snake and active contours proposed in~\citep{silveira2009comparison}, they have degraded with the change the pigments or the presence of hair. Furthermore, they failed to discriminate between the lesion and healthy skin, when the lesions have fuzzy boundaries. In turn, the clustering-based methods are not efficient with complex dermoscopic images.

\textbf{Deep learning-based approaches:} Recently, convolutional neural networks (CNNs) have been commonly applied for different tasks, such as image segmentation ~\citep{guo2018review}, object detection~\citep{zhao2019object} and image classification~\citep{rawat2017deep}. For image segmentation,  several CNNs are presented at the last five years. The fully convolutional network (FCN)~\citep{long2015fully} based model made the initial breakthrough, in which the encoder and decoder framework is employed for the segmentation task. Later on, different variations of the autoencoder networks were utilized for various segmentation tasks ~\citep{lateef2019survey}. The U-Net model prposed in~\citep{ronneberger2015u} outperformed state-of-the-art biomedical image segmentation using a handful amount of data. In order to memorize the features from first layers of the encoder and suppresses the singularities inherent in the loss of DCNNs The authors of~\citep{ronneberger2015u} also introduced a new concept called \textit{skip connection}, which features extracted by each encoder layer will be connected to the corresponding decoder layer .

Regarding skin lesion segmentation task, one of the well-known approaches proposed for skin lesion segmentation is the fully convolutional residual network (FCRN)~\citep{yu2017automated} that yields a detailed and accurate skin lesion segmentation by learning multiscale contextual features of the input image. Although FCRN gives good results, some factors, such as low contrast dermoscopic images, the presence of hairs and irregular lesion shapes degrade the segmentation results.  There are also several U-Net based models proposed for skin lesion segmentation task. For instance, a self-ensemble U-Net model proposed in~\citep{li2018semi} with a transformation that improved the effects of regularization by using the unlabeled data, achieving an Intersection over Union (IoU) score $79.87$\% on the ISBI 2017 test dataset. In~\citep{bissoto2018deep}, the authors proposed to use image processing tools to remove noise from the input dermoscopic images, and then the refined image fed to a U-Net based model to segment skin lesions, achieving an IoU score of $72.8\%$ on the ISIC2018 dataset. In~\citep{vesal2018skinnet}, a network called SkinNet has been proposed. Where the densely connected convolution layers are the core of the layers of SkinNet, which yielded an IoU score of $76.7$\% on the ISBI 2017 test dataset. In the SLSDeep model~\citep{sarker2018slsdeep}, a deep network has been proposed via using dilated residual convolution layers with a pyramid pooling network to extract contextual features from multiscale representations of a dermoscopic image. Besides, a combination of negative Log-likelihood and endpoint error functions has been used as a loss function to improve the boundaries of the segmented lesions. SLSDeep obtained an IoU score of $78.2$\% on the ISBI 2017 dataset. In turn, several FCN-based approaches are also applied to skin lesion segmentation. For example, a skin segmentation model based on the FCN architecture has been presented in~\citep {bi2019step} that with the ISBI 2017 dataset it provided an IoU score of $77.73$\%. Besides, a full-resolution convolutional network (FrCN) was introduced in~\citep{al2018skin} yielded an IoU score of $77.11$\% on the ISBI 2017 dataset. 
Regarding skin lesion segmentation, several methods based on GANs have been proposed. For instance, \citep{xue2018adversarial} has presented a GAN-based model that depened on ResNet blocks with skip connections. With the ISBI 2017 test dataset, it achieved an IoU score of $78.50$\%.  In~\citep{bisla2019skin}, a two-stream deep convolutional generative adversarial network with the ResNet-50 model, was proposed to jointly segment and classify skin lesions. In this study for removing the artifacts from the lesion images, the authors have used different pre-processing techniques. The work in \citep{bisla2019skin} has provided IoU scores of  $77.00$\% and $70.20$\%, respectively on ISBI2017 and ISIC2018 datasets.

\textbf{GAN-based approaches:}
In the literature, several lightweight models have been proposed for different applications, such as object detection and image segmentation. For instance,  in~\citep{sae2019convolutional}, DCNNs and the MobileNets have been used for skin image classification. The MobileNets model~\citep{howard2017mobilenets} was tested with use cases including object detection, fine-grain classification, face attributes, and large scale geo-localization. MobileNets is based on depth-wise separable convolutions. It has around 4 million parameters, while object detection models, such as YOLO-V3  and Mask-RCNN, have about 60  and 40 million parameters, respectively. These object detection models have been used in some skin lesion segmentation models to detect the region of the lesion or to provide initial segmentation results. In ~\citep{mishraalgoderm}, a two-step method was proposed for the analysis of skin lesions that accepts dermoscopic as well as cellphone images. Firstly, Mask-RCNN based on ResNet152 has been used to obtain an initial segmentation. Secondly, to refine the segmentation results, the initial segmentation has been fed to a superpixel segmentation method. The method proposed in \citep{mishraalgoderm} has provided accurate segmentation, however it is still a heavy model, since it exploits the ResNet152 backbone that has ~60.2 million parameters. Also, YOLO and the GrabCut algorithm have been combined in~\citep{unver2019skin} for skin lesion segmentation with a Dice score of 84.26 and a Jaccard score of 74.81 on ISBI  2017 dataset. However, these scores have considered much lower than the results of the state-of-the-art, such as SLSDeep that achieved a Dice score of 87.80 and a Jaccard score of 78.20. In turn, in \citep{paszke2016enet}, an efficient deep neural network named ENet for real-time segmentation with small number of parameters of ~1 million have been introduced. Enet has produced a Dice score of 82.70 and a Jaccard score of 74.10 with ISBI 2017.

\textbf{Attention mechanism-based approaches:} Recently, attention mechanisms have been used in DCNNs~\citep{mnih2014recurrent},~\citep{wang2017residual},~\citep{chen2016attention}. Attention mechanisms help DCNNs to pay more attention to the components with more enriched information. To that so,  using trainable layers, the current convolutional features are weighted by multiplying with learned soft weight vectors to produce channel-wise features, which yield useful features and suppressed the redundant ones.  As the attention mechanism focuses on relevant information in the collection of extracted features, it encourages DCNNs to learn faster with merely a small training dataset. In~\citep{chen2016attention}, an attention mechanism for softly weighting the multiscale features at each pixel location has been proposed. This attention mechanism is able to capture the key features at different scales and positions.  In~\citep{schlemper2018attention}, the use of a generalized self-gated soft-attention mechanism allows the convolutional layers to contextualize local features. This mechanism can be combined with existing deep learning segmentation or classification models while adding few trainable parameters. For medical image analysis,  \citep{oktay2018attention} has proposed an attention gate model that is able to determine the structures of lesions with different shapes and sizes. It can promote the task-relevant salient features and neglect irrelevant regions in the input images. Besides, \citep{mnih2014recurrent} has presented a visual attention mechanism, which can be cable of finding the salient regions and processing them at high resolution. 

Indeed, most of the skin lesion segmentation methods, as mentioned above, use profound models having a massive amount of parameters, which leads to reduce the potential impact of these methods on daily clinical practice. Besides, these heavy methods can not be used for real-time applications running on low resources embedded systems (e.g., mobiles). To address this issue, we propose MobileGAN, which is a lightweight and efficient GAN-based model. The number of parameters of any segmentation model depends on several factors, such as the number of convolutional layers used, image size of the input layer, fully connected layers. Unlike the related work, the MobileGAN model proposes a novel layer, \textit{factorized-attention module} (FCM), which combines two branches: residual 1-D factorized kernel convolution (factorized layer) and channel attention module. In the FCM, convolution is computed in a fashion that reduces the overall number of computational parameters. We employ a multiscale aggregation mechanism to extract skin lesions-relevant features at different scales to be able to segment of skin lesions of various shapes and sizes.  Besides, to encode contextual information into local features, both position attention module (PAM) and channel attention module (CAM) are utilized in order to accurately distinguish skin lesions from the healthy skin through the localized texture information. Notably, CAM and PAM can facilitate the training process of the model, since they encourage the model to learn skin-lesion relevant features. They also will not affect the number of the parameters of the training model.

\section{Methodology}
In this section, we describe in details the network architecture and its developed layers. In this work, our baseline network is GAN. 
The GAN, pix2pix, proposed in~\citep{isola2017image} has been applied to different medical applications, such as  medical image segmentation and classification. In general, two main network are the core of GAN, namely the generator $G$ and discriminator $D$. The generator consists of an autoencoder architecture (i.e., encoder and decoder networks), which can be trained to learn the mapping from an image from domain $A$ (dermoscopic images) to domain $B$ (segmented lesions). The discriminator is used to compare the generated segmentation masks with real segmented images. 
Figure~\ref{fig1:model1} presents the architecture of the proposed model, which has $G$ and $D$ networks as the pix2pix model. A multiscale block is used for aggregating the coarse-to-fine features of dermoscopic images to alleviate false detection due to the artifacts. The encoder and decoder parts of the generator and the discriminator networks are explained in detail in the next subsections. 

\subsubsection {The Encoder Network}
\begin{figure*}[!t]
	\centering
	\includegraphics[width=\textwidth]{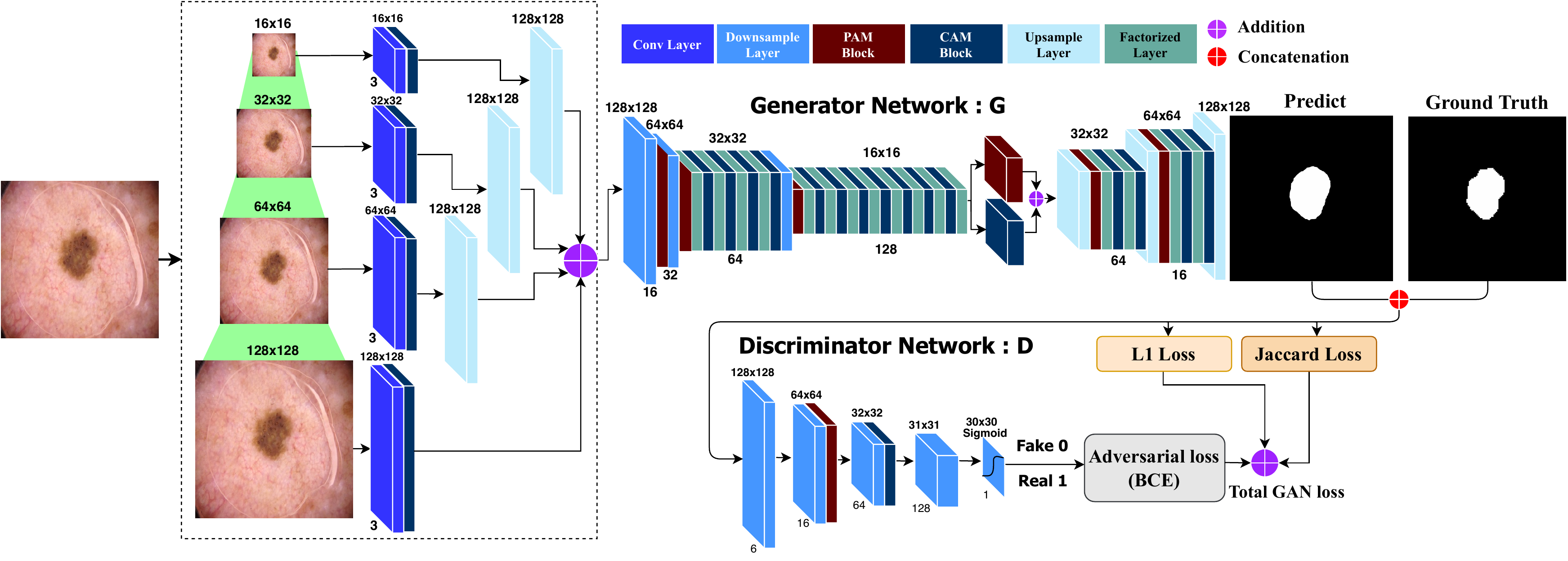}
	\caption{The architecture of MobileGAN: generator network (top) and discriminator network (bottom). }
	\label{fig1:model1}
\end{figure*}

Firstly, three scaled images are generated from the input images with the ratios of 1/8, 1/4, and 1/2 of the original size for the encoder of generator network $G$, as shown in Figure~\ref{fig1:model1}. This will help MobileGAN to be invariant to image resolution by deal with of object, and images in different resolutions and scales. These types of scale-invariant filters capture small pixels that can help to properly segment tiny skin lesions. In the multiscale block, four $3\times 3$ convolutional filters are then applied and followed by four CAMs to capture visual features dependencies in channel dimensions (more details are given below). The aggregated module is fed with four sizes of the features maps from up to down are $128 \times 128 \times 16$, $64 \times 64 \times 16$, $32 \times 32 \times 16$, and $16 \times 16 \times 16$, respectively. Afterward, the three lower-scale features are upsampled to the same size of the original input image by using the bilinear interpolation method and then average the four feature maps to generate a $128 \times 128 \times 16$ feature map. 

The aforementioned multiscale strategy helps the encoder to extract low level features in different scales to cope with shadows. Besides, the resulted feature maps are created in both spatial and channel domains. The resulted 16 feature maps are fed into two downsampling-attention layers. Each downsampling-attention layers comprises a convolutional layer followed by a max-pooling of 2, and then a position attention module (PAM) to capture the spatial features. The two layers produce 64 feature maps that are fed into the next four factorized-attention module (FCM).  Each FCM includes a 1-D kernel factorized layer followed by a CAM. A downsampling-attention layer feeds by the resulting feature maps to produce $128$ feature maps that are fed into the next eight FCM. The result of the eighth FCM is fed into a 1-D kernel factorized layer followed by two parallel attention blocks: one for CAM and the other for PAM that is summed to capture visual features independently to position and channel dimensions. In the end of the encoder, the final 128 feature maps are fed into the decoder to construct the segmented image determining the boundaries of skin lesions.

\begin{figure}[!t]
\centering
\includegraphics[width=0.7\textwidth]{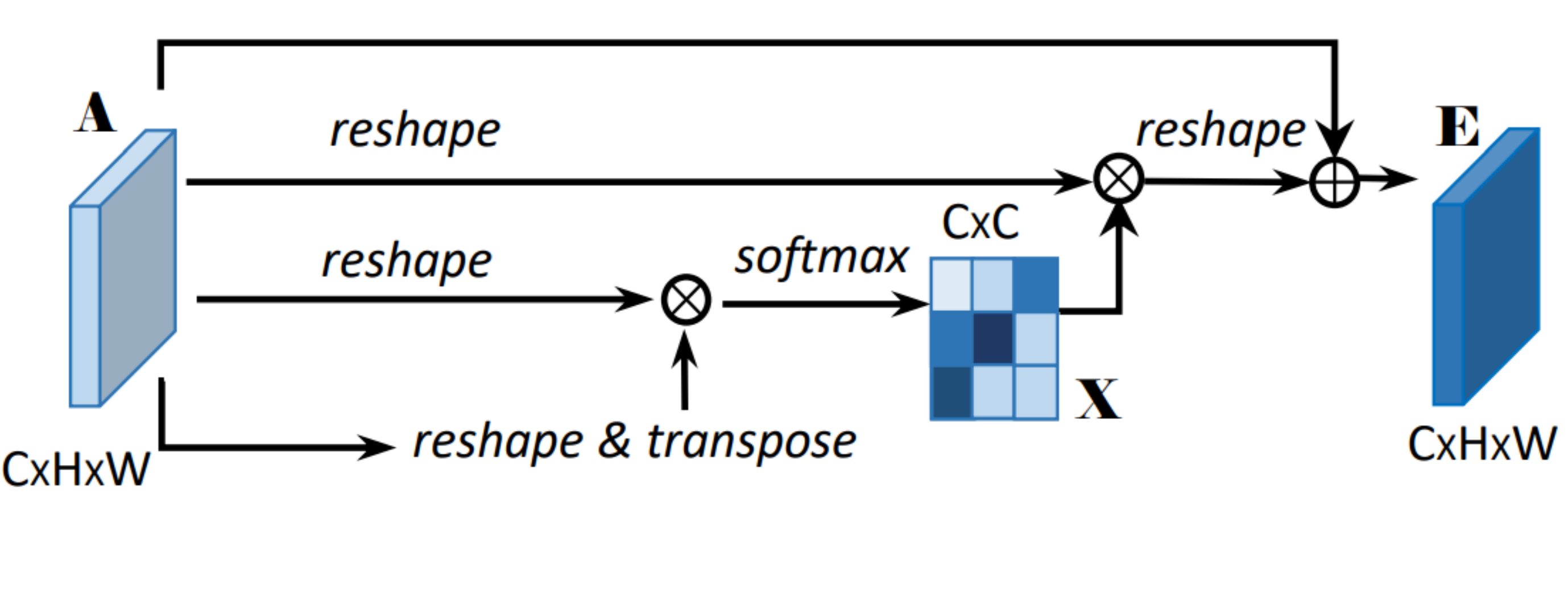}
\caption{Channel attention module (CAM) architecture.}
\label{fig:cam}
\end{figure}

\textbf{Channel attention module (CAM):} The feature maps consists of a set of channels that each one can be noted as a class-specific response representing high-level features. However many semantic responses (i.e. channels) are correlated with each other. Consequently, in this module by Using the inter-dependencies among channel maps, we can highlight inter-dependent feature maps and update the feature representation of specific interpretation. Thus, a channel attention module is built to model inter-dependencies among channels explicitly. Figure~\ref{fig:cam} shows the composition of the channel attention module. The channel attention map $\mathbf{X} \in \mathbb{R}^{C \times C}$ calculated from the original features $\mathbf{A} \in \mathbb{R}^{C \times H \times W}$ directly calculates from the position attention module, where $C$, $H$ and $W$ are channels, height and width of the input image, respectively . Clearly, $\mathbf{A}$ is reshaped to $ \mathbb{R}^{C \times N} $, where $N=H \times W$ is the number of features. A matrix multiplication between $\mathbf{A}$ and the transpose of $\mathbf{A}$ is then performed. Finally in order to generate the channel attention map $\mathbf{X} \in \mathbb{R}^{C \times C}$, a softmax function is applied as follows:
\begin{equation}
x_{ji}=\frac{exp(A_{i}\cdot A_{j})}{\sum_{i=1}^Cexp(A_{i}\cdot A_{j})},
\end{equation}

where $x_{ji}$ calculates the ${i^{th}}$ channel impact on the ${j^{th}}$ channel. Besides, matrix multiplication is performed between the transpose of $\mathbf{X}$ and $\mathbf{A}$. The result of the multiplication is reshaped to $\mathbb{R}^{C \times H \times W}$ that is then multiplied by a scale parameter $ \gamma$ and performed an element-wise addition operation with $\mathbf{A}$ to get the final output $\mathbf{E}\in \mathbb{R}^{C \times H \times W}$ as follows:
\begin{equation}
E_{j} = \gamma \sum_{i=1}^C(x_{ji}A_{i} ) + A_{j},
\label{equ2}
\end{equation}

where $\gamma$ continuously learns a weight from 0. The final feature of each channel is a weighted sum of the features of all channels and original features, which models the long-range semantic dependencies between feature maps as shown in the Equation~\ref{equ2}. Thus, it is obvious that CAM is able highlight class-dependent feature maps and discriminatively supports a feature boost that can be not produced by the convolution layers. 

\begin{figure}[!t]
\centering
\includegraphics[width=0.7\textwidth]{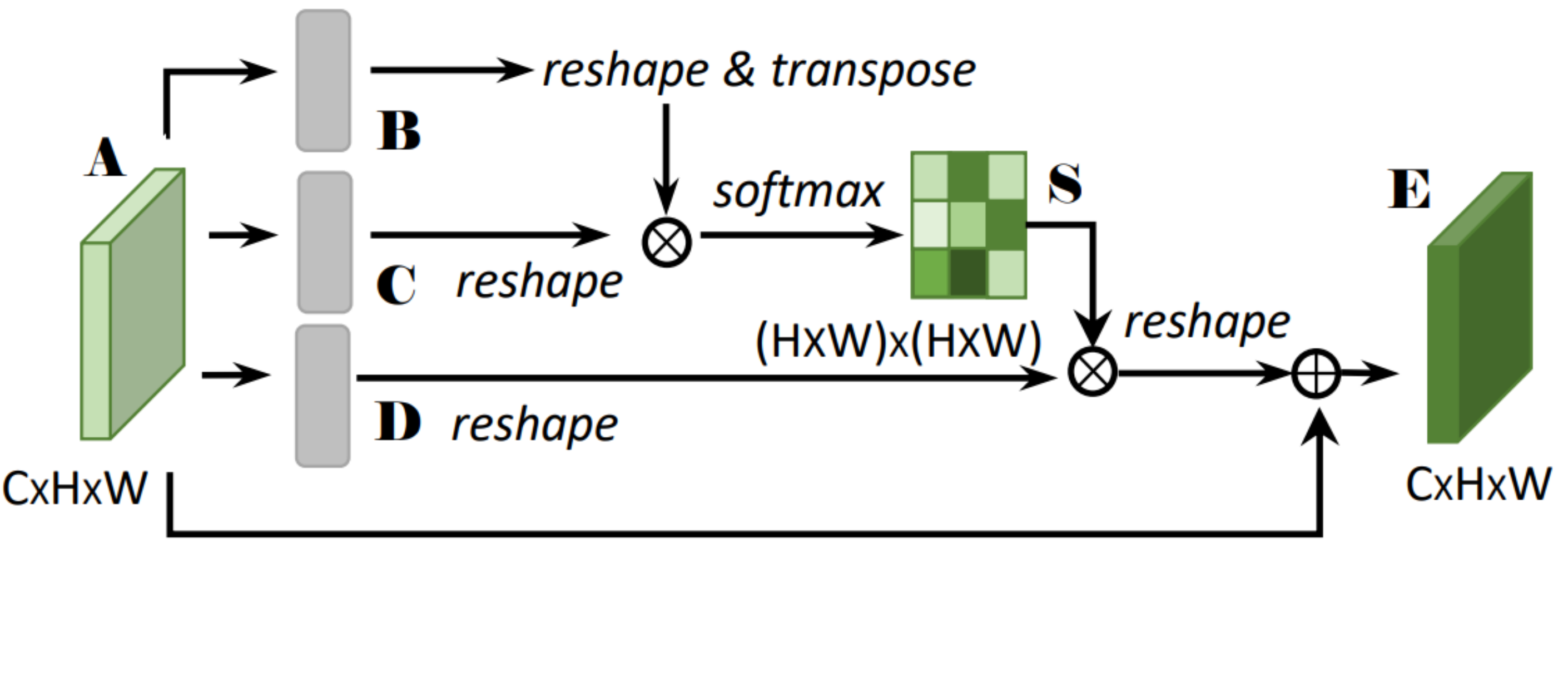}
\caption{Position attention module (PAM) architecture.}
\label{fig:pam}
\end{figure}

\textbf{Position attention module (PAM):}
To get accurate skin lesion segmentation, discriminant feature representations are essential that can be achieved by capturing long-range contextual information. Thus, in MobileGAN, we exploit a position attention module to model strong contextual links over local feature descriptions. A comprehensive series of contextual information into local features can be encoded by the position attention module. Ultimately, the spatial context is refined by aggregating the spatial features.
Given a local feature $\mathbf{A} \in \mathbb{R}^{C \times H \times W}$ that is fed to into a convolution layers with batch normalization and ReLU to produce two new feature maps $\mathbf{B}$ and $\mathbf{C}$, respectively, where $\{\mathbf{B},\mathbf{C}\} \in \mathbb{R}^{C \times H \times W}$ shown in Figure~\ref{fig:pam}. We then resize them to $ \mathbb{R}^{C \times N} $, where $N=H \times W$ is the number of features. Finally, a matrix multiplication between the transpose of $\mathbf{C}$ and $\mathbf{B}$ is performed, and to estimate the spatial attention map $\mathbf{S} \in \mathbb{R}^{N \times N}$ a softmax function $S$ is applied:
\begin{equation}
s_{ji}=\frac{exp(B_{i}\cdot C_{j})}{\sum_{i=1}^Nexp(B_{i}\cdot C_{j})},
\end{equation}

where $s_{ji}$ means the ${i^{th}}$ position's contact on ${j^{th}}$ position. Consequently, we can say that a softmax function $S$ attempts to find the correlation between two spatial positions in the input feature maps. Afterwards, the feature $\mathbf{A} $ is fed into a convolutional layer with batch normalization and ReLU to create a new feature map $\mathbf{D} \in \mathbb{R}^{C \times H \times W}$ that is resized to $ \mathbb{R}^{C \times N}$.
Matrix multiplication between $\mathbf{D}$ and the transpose of $\mathbf{S}$ are then performed to have new feature maps belonging $\mathbb{R}^{C \times H \times W}$. Eventually, a multiplication operation by a scale parameter $ \eta $ and an element-wise addition operation are performed with the features $\mathbf{A}$ to get the final output $\mathbf{E}\in \mathbb{R}^{C \times H \times W}$ as follows:
\begin{equation}
E_{j} = \eta \sum_{i=1}^N(s_{ji}D_{i} ) + A_{j},
\label{equ1}%
\end{equation}

where $\eta$ is initialized as $0$ as defined in~\citep{zhang2018self}. Based on (\ref{equ1}), the resulting feature $\mathbf{E}$ at every position is a weighted sum of the features of the complete neighbours of original features. Thus, it is obvious that the position attention module can yield a global contextual representation and selectively aggregate the context according to a spatial attention map by generating related semantic features that can achieve mutual gains and improve intra-class semantic consistency.

\begin{figure}[!t]
	\centering
	\includegraphics[width=0.7\textwidth]{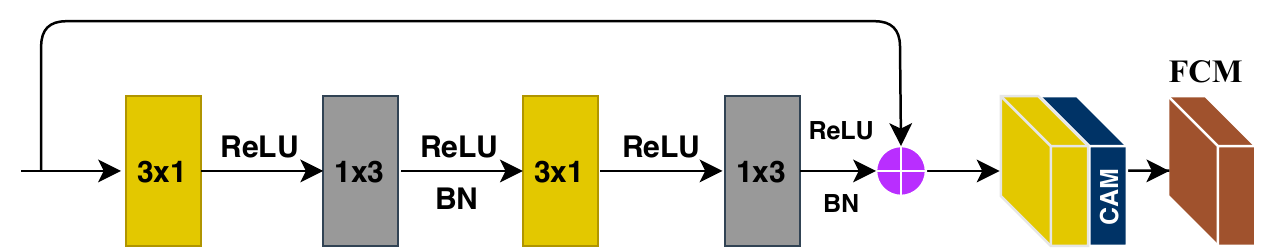}
	\caption{1-D Factorized-attention module (FCM) architecture. }
	\label{fig3:factor}
\end{figure}

\textbf{Factorized-attention module (FCM):}
For reducing the computation complexity, MobileGAN use residual 1-D kernel factorized layer. We assume that the weights of a typical 2D convolutional layer is indicated by $\textbf{W}\in\mathbb{R}^{C\times d^h\times d^v\times F}$, where $C$ is the number of input planes, $F$ is the number of output planes (feature maps) and $d^h\times d^v$ indicates to the kernel size of every feature map (typically $d^h \equiv d^v\equiv d$). Let $b\in\mathbb{R}^{F}$ be the vector denoting the bias term for every filter and $\textbf{f}^\textbf{i} \in\mathbb{R}^{d^{h} \times d^{v}}$ indicates the $i^{th}$ kernel of a layer. Thus, the rank-1 constraint can be rewritten, $\textbf{f}^\textbf{i}$, as a linear combination of 1D filters:
\begin{equation}
\centering
\textbf{f}^\textbf{i}=\sum_{k=1}^{K} \sigma_{k}^{i} \Bar{v}_{k}^{i}\big( \Bar{h}_{k}^{i}\big)^T,
\end{equation}

where $\sigma_{k}^{i}$ is a scalar weight and $K$ is a rank of $\textbf{f}^\textbf{i}$, in turn the length of vectors, $\Bar{v}_{k}^{i}$ and $\big(\Bar{h}_{k}^{i}\big)^T$ is $d$. Thus, the $i{th}$ output of the decomposed layer, $a_{i}^{1}$, is expressed as a function of its input $a_{*}^{0}$:
\begin {equation}
a_{i}^{1} = \varphi\bigg( b_{i}^{h} + \sum_{l=1}^{L} \Bar{h}_{il}^{T} * \bigg[\varphi\bigg(b_{l}^{v} + \sum_{c=1}^{C} \Bar{v}_{lc} * a_{c}^{0} \bigg)\bigg]\bigg),
\end{equation}

where the function, $\varphi(.)$, can be represented by non-linearity of the 1D decomposed filters, which can be implemented with ReLU. Finally, to get the final representation of the FCM, the output feature maps from 1-D kernel factorized layer is fed to the channel attention module. the architecture of the factorized-attention module used in MobileGAN is shown in Figure~\ref{fig3:factor}.

\subsubsection{The Decoder Network} The final output from the last encoding layer fed into a upsample-attention layer which consists of a deconvolutional layer with PAM block. Afterwards, the resulted output fed into two consecutive FCM. Again, another upsample-attention layer with two sequential FCM are used to achieved the final feature maps. Finally, the final feature maps is upsampled in order to obtain the segmented image. In order to convert the output to binary masks, a threshold of 0.5 is utilized. In all layers of both encoder and decoder networks, convolutional and deconvolutional filters with a kernel size of $3\times 3$, and a stride of 2 with a padding of $1$ are used. 

\subsubsection{The Discriminator Network} It includes four downsampling layers. The First three layers are convolutional layers with a kernel size of $4\times 4$, a stride of 2, and a padding of 1. Besides, a PAM block is used to the second downsampling layer, while in the third layer, a CAM block is added. However, in the final layer,  a sigmoid activation function is utilized to discriminate the generated binary mask versus real one.

\subsection{Model training}
In MobileGAN, back-propagation in an adversarial fashion is used for alternately training the $G$ and $D$ networks. Firstly, by using the gradients computed from the loss function and with fixing $G$, we train the $D$ network for one time. We then fix the $D$ network and train the $G$ network for another time by using the gradients computed from the same loss function passed from $D$ to $G$. 
Assume a skin lesion image is $x$, and the ground-truth of the segmented image is $y$. Let $z$ is a random variable that can be introduced as a dropout in the layers of the decoder, which helps to avoid overfitting of the model and generalize the learning process. Thus, the outputs of the generator and the discriminator can be expressed as $G(x, z)$ and $D(x, G(x,z))$, respectively. Consequently, we can define the loss function of the generator $G$ including three main error functions: binary cross entropy loss, $L_1$ norm to reduce the outliers, and Jaccard loss to increase the intersection between the segmented images and the ground-truth images:
\begin{equation}
\begin{split}
\ell_{Gen}(G, D) & = \mathbb{E}_{x,y,z}(-\log(D(x, G(x,z)))) \\
& + \lambda \mathbb{E}_{x,y,z}(\ell_{L_1}(y, G(x,z))) \\
& + \alpha \mathbb{E}_{x,y,z}(\ell_{Jaccard\:loss}(y, G(x,z))),
\end{split}
\end{equation}

where $\lambda$ and $\alpha$ are empirical weighting factors. In many cases, the adversarial loss term yields too slow learning, thus MobileGAN uses the \textit{$L_1$ loss} for boosting the learning process by properly formulating the gradient towards the expected segmented lesion boundaries. 
In addition, we consider the optimization of the \textit{Jaccard loss}  for the lesion classes. Let $Gt$ be the hand drawn ground truth of the lesion region, and $Pd$ its respective computer-generated segmentation mask, then the binary Jaccard loss  is based on the Jaccard distance defined as follows~\citep{yuan2017automatic}:
\begin{equation}
d_J(Gt,Pd)= 1-J(Gt,Pd) = 1- \frac{( Gt \cap Pd)}{|Gt|+|Pd|-|Gt\cap Pd|}.
\end{equation}

A non-differentiable function ${d_J(Gt,Pd)}$ can be introduced for loss minimization; however, it is not easy to directly apply such function for back-propagation. In order to generate a binary mask from continuous MobileGAN output for each iteration during optimization and to reduce the computation cost, we use the \textit{Jaccard loss} function that can be defined as:
\begin{equation}
L_{d_J}= 1- \frac{\sum_{x,y}(g_{xy} , p_{xy})}{\sum_{x,y}g^2_{xy}+\sum_{x,y}p^2_{xy}-\sum_{x,y}(g_{xy} p_{xy})},
\end{equation}

where $g_{xy}$ and $p_{xy}$ are the pixel values at $(x,y)$ in a ground-truth and predicted mask, respectively. To balance the pixels of lesion regions and background, a weight map is used; however, it is not the case for the  defined \textit{Jaccard loss}, since the Jaccard loss function is differentiable:
\begin{equation}
\begin{split}
JL= \frac{\delta L_{d_J}}{\delta L_{p_{xy}}} = -\frac{g_{xy}[ \sum_{x,y}g^2_{xy}+\sum_{x,y}p^2_{xy}-\sum_{x,y}(g_{xy} p_{xy})]}{[\sum_{x,y}g^2_{xy}+\sum_{x,y}p^2_{xy}-\sum_{x,y}(g_{xy} p_{xy})]^2} \\ 
+ \frac{(2p_{x,y}-g_{xy})[\sum_{x,y}(g_{xy} p_{xy})]}{[\sum_{x,y}g^2_{xy}+\sum_{x,y}p^2_{xy}-\sum_{x,y}(g_{xy} p_{xy})]^2}.
\end{split}
\end{equation}

During network training, the \textit{Jaccard loss} can be efficiently integrated into the back-propagation. If the generator network is optimized correctly, the values of $D(x, G(x,z))$ approach $1.0$, which means that the discriminator cannot differentiate the generated segmentation mask from the ground truth. In such a case, $L_1$ and Jaccard losses should approach $0.0$, indicating that each generated mask matches the corresponding ground truth mask both in overall pixel-to-pixel distances ($L_1$) and in tight convex surrogate (Jaccard loss) to all intersection-over-union (IoU). Thus, the loss function of the discriminator $D$ can be defined as follows:
\begin{equation}
\centering
\label{Ldis}
\begin{split}
\ell_{Dis}(G, D) &= \mathbb{E}_{x,y,z}(-\log(D(x, y))) \\ 
& +\mathbb{E}_{x,y,z}(-\log(1-D(x, G(x,z)))).
\end{split}
\end{equation}

In (\ref{Ldis}), There are two terms to compute the binary cross-entropy (BCE) loss using two images:  the term $-\log(D(x, y))$ for ground truth images, and the other one $-\log(1-D(x, G(x, z))$ for the predicted image. The optimizer fits $D$ by maximizing the loss values for the ground-truth images and minimizing the loss values for the predicted images. We assume that the expected class for ground truth and generated images are $1$ and $0$, respectively

\section{Experiments}

\textbf{Datasets:} The efficacy of the proposed model, MobileGAN, is assessed on two publicly available datasets of dermoscopic images for skin lesion analysis: IEEE International Symposium on Biomedical Imaging, (ISBI 2017) and Skin Lesion Analysis Towards Melanoma Detection, grand challenge datasets (ISIC 2018)~\footnote{https://challenge.isic-archive.com}. ISBI 2017 dataset was divided into training, validation, and testing sets with 2000, 150, and 600 images, respectively. In turn, the ISIC 2018 dataset includes 2,594 images with the corresponding ground truth masks annotated by expert dermatologists. The validation and testing sets contain 100 and 1,000 images, respectively, without ground truth (evaluated on the  ISIC2018 validation leaderboard~\footnote{https://challenge.isic-archive.com/} only). In our experiments, we used 80\% of the training set of the ISIC 2018 dataset for training the segmentation models and 20\% for validation, as proposed in~\citep{al2018skin}. Note that we trained, validated, and tested MobileGAN individually on the ISBI 2017 and ISIC 2018 datasets.  

\textbf{Evaluation metrics:} In this paper, we use five evaluation metrics to evaluate the performance of MobileGAN. With the ISBI 2017 dataset, we use Jaccard similarity coefficient (JSC), Dice similarity coefficient (DSC), accuracy (ACC), specificity (SPE) and sensitivity (SEN)~\cite{ISIC}. For both the ground-truth $y$ and the predicted image $x$, the true positive (TP) rate can be defined as $TP=y\cap x$, which is the area of the segmented region common in both $x$ and $y$. The false positive (FP) rate can be defined as $FP$ = $\overline{y} \cap x$, which is the segmented area not belonging to $y$. The false-negative (FN) rate is defined as $FN$ = $y \cap \overline{x}$, which is the actual area missed in the predicated image. The true negative (TN) set can be defined as $TN$  = $\overline{y}\cap\overline{x}$, which is the set of image background common in both $x$ and $y$. Below, we present the mathematical expressions of the five metrics: ACC, DSC, JSC, SEN and SPE. 
\begin{equation}
ACC = \frac{TP+TN}{TP+TN+FP+FN} 
\end{equation}
\begin{equation}
DSC = \frac{2.TP}{2.TP+FP+FN}
\end{equation}
\begin{equation}
JSC = \frac{TP}{TP+FP+FN}
\end{equation}
\begin{equation}
SEN = \frac{TP}{TP+FN}
\end{equation}
\begin{equation}
SPE = \frac{TN}{TN+FP}
\end{equation}
The predicted lesion masks of the ISIC 2018 challenge are assessed using a threshold JSC ($JSC_{th}$)~\citep{codella2019skin}. Through comparing each pixel of the predicted image to its corresponding pixel in the ground-truth mask, the JSC for each test image is calculated. Thus, the $JSC_{th}$ can be computed as follows:
\begin{equation}
JSC_{th}=
\begin{cases}
JSC & \text{if}\ JSC \gneq 0.65, \\
0 & {otherwise}.
\end{cases}
\label{eq:logloss1}
\end{equation} 
where the images with $JSC<0.65$ will be given a score of 0.

\textbf{Data augmentation:} We augment the two datasets by flipping the images horizontally and vertically, applying gamma reconstruction, and changing the contrast using adaptive histogram equalization (CLAHE) with different values on the original RGB images. We increased the total number of training images to 16000 and 16600 after applying the data augmentation on ISBI 2017 and ISIC 2018 train dataset.

\textbf{Implementation:} To train the network, our experiments were carried on NVIDIA 1080Ti with 11GB memory taking ~8 hour. We implemented The proposed model on the PyTorch framework~\footnote{https://pytorch.org/}. Besides, the Adam optimizer~\citep{kingma2014adam} with the parameters $\beta_1 = 0.5$ and $\beta_2 = 0.999$ was used. We set the learning rate to $0.0002$ and the batch size is set to $8$. We also set the weighting factors of $L_1$-norm loss and Jaccard loss ($\lambda$ and $\alpha$) to 0.1 and 0.5, respectively. 

\textbf{Experimental results:} For ISBI 2017 and ISIC 2018, the size of the images ranges from $542\times 718$ to $2848\times 4288$ considered very large to train the proposed model. Thus, to speed up the training process, we resized the input images to $H\times W$ pixels, where $H$ and $W$ are height and width of the image fed to the network. For select the best of the image size yielding the best accuracy, we trained and tested our model with three image sizes ($64\times 64$, $128\times 128$ and $256\times 256$). The best segmentation results are obtained with the input size of $128\times 128$ (ablation study is given below). 

\begin{table}[!t]
	\centering
	\caption{A comparison between the proposed model and 6 skin lesion segmentation methods on the ISBI 2017 dataset (test set) in terms of the accuracy and number of parameters in million (Params(M)).}
	\label{table2_eval2017test}
	\scalebox{0.75}{
		\begin{tabular}{@{}lllllll@{}}
			\hline \hline
			\quad  {{\textbf{Methods}}} \quad&\quad  {{\textbf{ACC}}} \quad&\quad  {{\textbf{DSC}}} \quad&\quad  {{\textbf{JSC}}} \quad&\quad  {{\textbf{SEN}}} \quad&\quad  {{\textbf{SPE}}}
			\quad&  {{\textbf{Params(M)}}}
			\\ \hline \hline  
			\quad FCN~\citep{long2015fully} \quad&\quad 92.72 \quad&\quad 83.83 \quad&\quad 72.17 \quad&\quad 79.98 \quad&\quad 96.66 \quad&\quad 134.3 \\ 
			\quad U-Net~\citep{ronneberger2015u} \quad&\quad 90.14 \quad&\quad 76.27 \quad&\quad 61.64 \quad&\quad 67.15 \quad&\quad 97.24 \quad&\quad 12.3 \\ 
			\quad SegNet~\citep{badrinarayanan2017segnet} \quad&\quad 91.76 \quad&\quad 82.09 \quad&\quad 69.63 \quad&\quad 80.05 \quad&\quad 95.37 \quad&\quad 11.50 \\ 
			\quad FrCN~\citep{al2018skin} \quad&\quad 94.03 \quad&\quad 87.08 \quad&\quad 77.11 \quad&\quad {85.40} \quad&\quad 96.69 \quad&\quad 16.30 \\ 
			\quad SLSDeep~\citep{sarker2018slsdeep} \quad&\quad 93.60 \quad&\quad 87.80 \quad&\quad 78.2 \quad&\quad 81.60 \quad&\quad 98.30 \quad&\quad 46.65 \\ 
			\quad SegAN~\citep{xue2018adversarial} \quad&\quad 94.10 \quad&\quad 86.70 &\quad 78.50 \quad&\quad - \quad&\quad - \quad&\quad 382.17 \\ 
			
			\quad YOLO+grabcut~\citep{unver2019skin} \quad&\quad 93.39 \quad&\quad 84.26 & \quad 74.81 \quad&\quad \textbf{90.82} \quad&\quad 92.68 \quad&\quad - \\ 
			
			\quad ENet~\citep{paszke2016enet} \quad&\quad 92.0 \quad&\quad 82.7 & \quad 74.1 \quad&\quad - \quad&\quad - \quad&\quad \textbf{0.36} \\ 
			
			\quad \textbf{Proposed MobileGAN} \quad&\quad \textbf{97.61} \quad&\quad \textbf{90.63} \quad&\quad \textbf{81.98} \quad&\quad 87.81 \quad&\quad \textbf{99.92}\quad&\quad  2.35 \\ \hline \hline 
		\end{tabular}%
	}
\end{table}

In Table~\ref{table2_eval2017test} and Table~\ref{table3_eval2018val}, quantitative results of the proposed model on ISBI 2017 test and ISIC 2018 validation sets are shown. In Table~\ref{table2_eval2017test}, we compared the MobileGAN with eight skin lesion segmentation methods: FCN ~\citep{long2015fully}, U-Net~\citep{ronneberger2015u}, SegNet~\citep{badrinarayanan2017segnet}, FrCN~\citep{al2018skin}, SLSDeep~\citep{sarker2018slsdeep}, SegAN~\citep{xue2018adversarial}, YOLO+grabcut~\citep{unver2019skin} and ENet~\citep{paszke2016enet} using ISBI 2017 test dataset. We took all the test results of FCN, U-Net, SegNet, FrCN from~\citep{al2018skin} that used the same dataset. Table ~\ref{table2_eval2017test} shows that MobileGAN can outperform the all tested methods in terms of ACC, DSC, JSC, and SPE metrics. MobileGAN yields the ACC and JSC of $97.61\%$ and $81.98\%$, which is $3.51\%$ and $3.48\%$ higher than the ACC and JSC of the second-best (segAN) method. Similarly, it yields the DSC and SPE of $90.63\%$ and $99.92\%$, which is $2.83\%$ and $1.62\%$ higher than the DSC and SPE of the second-best (SLSDeep) method. In turn, the YOLO+grabcut yields the SEN of $90.82\%$, which is $3.01\%$ higher than the MobileGAN. MobileGAN has a small number of parameters (2.35 million parameters) compared to all other methods except ENet. Although ENet is a lightweight model (0.36 million parameters), its segmentation results are worse than MobileGAN.

\begin{table}[!b]
	\centering
	\caption{Evaluating the proposed MobileGAN model on the ISIC 2018 validation dataset in terms of the $JSC_{th}$ and number of parameters in million (Params(M)).}
	\label{table3_eval2018val}
	\scalebox{0.75}{
		\begin{tabular}{@{}lllll@{}}
			\hline \hline
			\quad  {\textbf{Methods}} &\quad  {{\textbf{$JSC_{th}$}}} & \quad  {\textbf{Params(M)}}\\ \hline \hline  
			\quad FCN~\citep{long2015fully} &\quad  {74.70} &\quad \quad {134.30} \\
			\quad U-Net~\citep{ronneberger2015u} &\quad {54.40} &\quad \quad {12.30} \\ 
			\quad SegNet~\citep{badrinarayanan2017segnet} &\quad {69.50} &\quad \quad { 11.50} \\ 
			\quad FrCN~\citep{al2018skin} &\quad {74.60} &\quad \quad { 16.30} \\ 
			\quad GAN-FCN~\citep{bi2018improving} &\quad {77.80} &\quad \quad { 10.61} \\ 
			\quad {Rcnn-superpixels}~\citep{mishraalgoderm} &\quad {\textbf{83.00}}&\quad \quad {-}\\ 
			\quad {Mask R-CNN}~\citep{sivanesan2019unsupervised} &\quad {78.80}&\quad \quad {-}\\ 
			\quad \textbf{Proposed MobileGAN} &\quad {78.40}&\quad  \quad {\textbf{2.35 }}\\ \hline \hline 
		\end{tabular}%
	}
\end{table}

Regarding the ISIC 2018 validation dataset, we compared MobileGAN with the FCN, U-Net, SegNet, FrCN, GAN-FCN, Rcnn-superpixels, Mask R-CNN models, as shown in Table~\ref{table3_eval2018val}. Note that the ISIC 2018 dataset includes 100 images for validation and 1,000 images for testing, without ground truth. However, the evaluation is only done on the  ISIC 2018 validation leaderboard. We used the validation evaluation of FCN, U-Net, SegNet, FrCN from the literature~\citep{alautomatic}. The proposed MobileGAN model achieves the highest $JSC_{th}$ score compared to the GAN-FCN and FrCN with an improvement of $0.6\%$ and $3.8\%$, respectively. Notably, MobileGAN has a number of parameters much lower than all compared models. 

\begin{figure*}[!t]
	\centering
	\includegraphics[width=\textwidth]{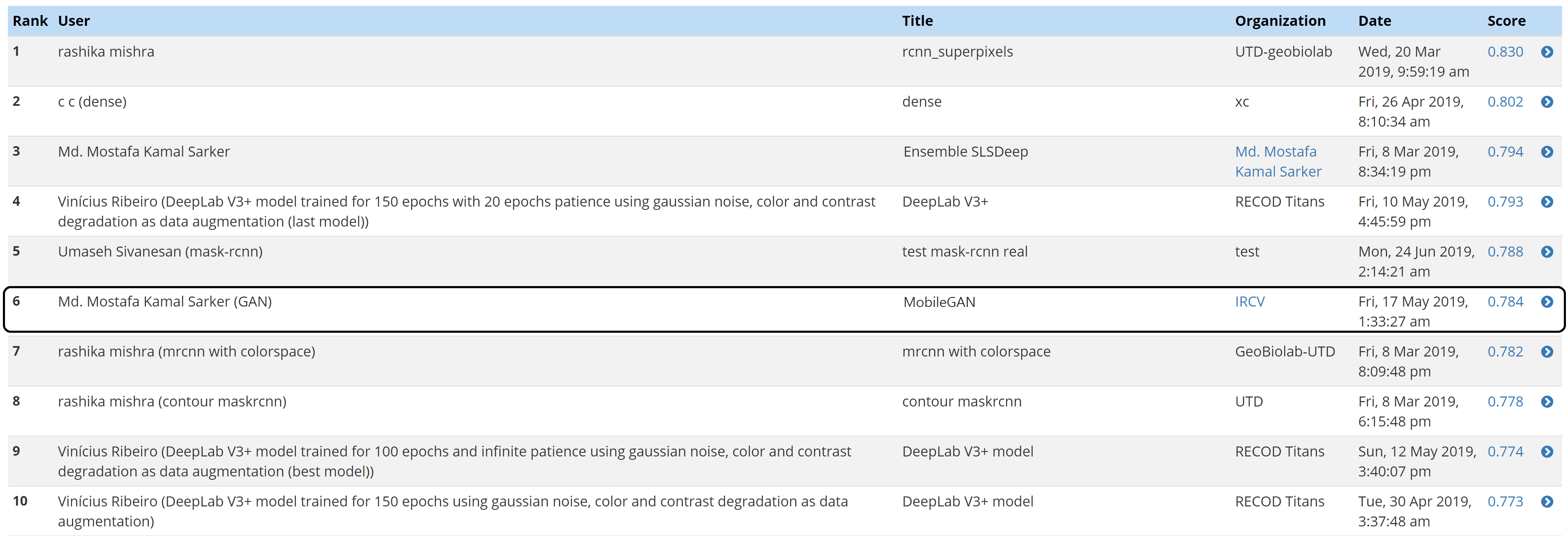}
	\caption{The rank of MobileGAN on the ISIC 2018 leaderboard challenge (screenshot). The proposed MobileGAN is highlighted.}
	\label{fig_lb:fig_lb}
\end{figure*}

In addition, we compared our model, MobileGAN, with the six segmentation models (i.e., the FCN, U-Net, SegNet, FrCN, SegAN, and GAN-FCN) in terms of the number of the parameters. The MobileGAN model has only 2.35 million parameters, while the GAN-FCN model (the closest one) has 10.61 million parameters. In turn, the SegAN model that is based on the traditional GAN model is the heaviest model with 382.17 million parameters.  We emphasize that the exploitation of 1-D kernel, PAM and CAM significantly reduces the number of parameters of the proposed model, MobileGAN, which  has 57x, 5x, 4x, 6x, and 19x times lower parameters than the FCN, U-Net, SegNet, FrCN, and SLSDeep models, respectively. The rank of MobileGAN with the validation set of the ISIC 2018 challenge is shown Figure~\ref{fig_lb:fig_lb}. Note that the proposed model is highlighted, and has the title MobileGAN and submitted by IRCV on 17th of May 2019. It has been ranked on the $6th$ position at the time of submission. The models preceding MobileGAN on the leaderboard (rank 1 to 5) are using different networks; however, their baseline models are the same as residual networks (i.e., ResNet)~\citep{he2016deep}, noting that the number of parameters of ResNet-50 is 23 million. In other words, the number of parameters of these models is higher than MobileGAN. In turn, Figure~\ref{fig_lb:fig_lb} and Table~\ref{table3_eval2018val} show that the Mask R-CNN~\citep{sivanesan2019unsupervised} and Rcnn-superpixels~\citep{mishraalgoderm} models achieve $JSC_{th}$ a bit higher than MobileGAN, however the authors of Mask R-CNN~\citep{sivanesan2019unsupervised} and  Rcnn-superpixels~\citep{mishraalgoderm} models did not mention the number of parameters of each model. As shown, both Mask R-CNN~\citep{sivanesan2019unsupervised} and Rcnn-superpixels~\citep{mishraalgoderm} exploit the ResNet 101 backbone. Since the number of parameters of ResNet 101 is  ~44.5 million, and the Mask  R-CNN and Rcnn-superpixels models are much heavier than MobileGAN. As shown, MobileGAN is a lightweight model and achieves competitive skin segmentation results when compared to state-of-the-art segmentation models.

\begin{figure}[!t]
	\centering
	\includegraphics[trim={0 0 0 0},clip,scale=.45]{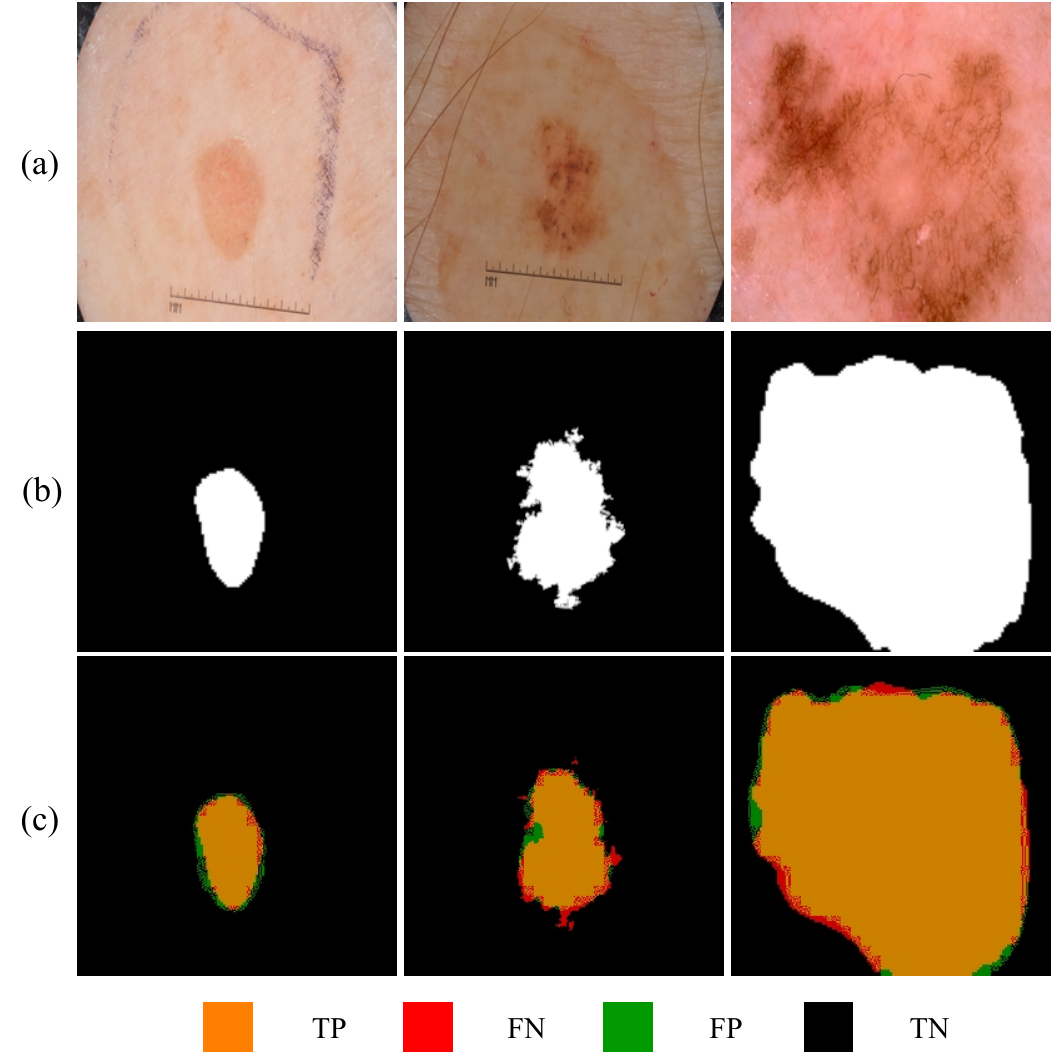}
	\includegraphics[trim={0 0 0 0},clip, scale=.45]{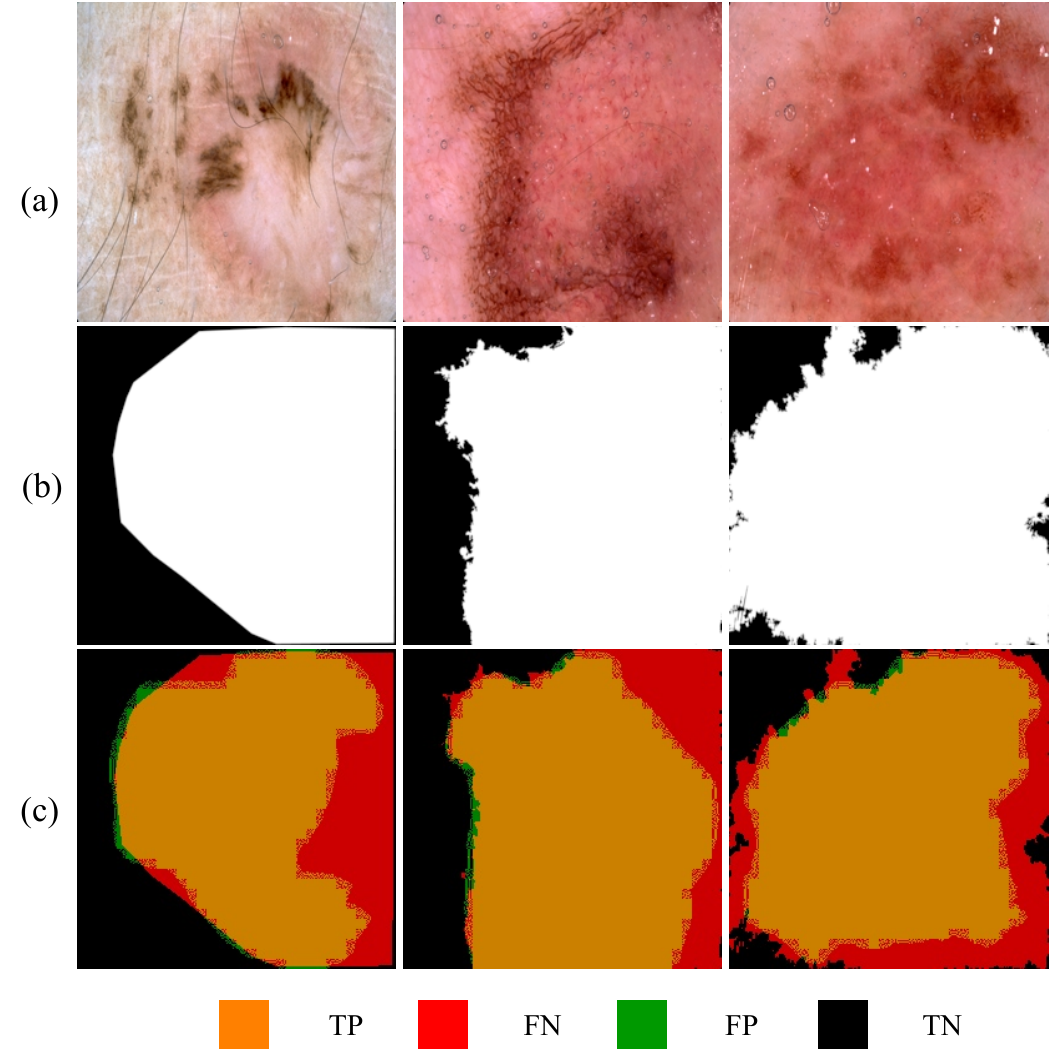}
	\caption{Segmentation results of MobileGAN: (a) input image (b) ground truth (c) \textbf{left:} accurately segmented lesions (c) \textbf{right:} incorrectly segmented lesions}
	\label{fig2:results}
\end{figure}

Qualitative segmentation results of the MobileGAN model with some examples of the ISBI 2017 test dataset are shown in Figure~\ref{fig2:results}. As shown in Figure~\ref{fig2:results} (left), although the tested images have a high similarity between the color of the lesion and the skin regions, fuzzy boundaries and even tiny lesions, the MobileGAN model accurately segments the boundary of each skin lesion with an accuracy of about $95\%$.  In Figure~\ref{fig2:results} (right), the shown examples have tiny skin regions (i.e., the background) compared to lesion regions. The lesion regions fill most of the image and intersect three margins of the images. In these cases, MobileGAN yields inaccurate segmentation because it is a bit difficult to segment the boundaries of skin lesions accurately when there is no proper boundary between the lesion and healthy skin tissue. In such cases, it is hard for any segmentation model to precisely delineate the shape of the lesion region to get an acceptable segmentation.

\begin{table}[!t]
	\centering
	\caption{Comparison between the inference times of MobileGAN (a lightweight architecture) and GAN-FCN  at different image resolutions.}
	\label{table4_inference}
	\scalebox{0.72}{
		\begin{tabular}{@{}llllllllll@{}}		
			\hline \hline
			\multirow{2}{*}{ \textbf{Model}} &&{{\textbf{64x64}}} &&{{\textbf{128x128}}} &&{{\textbf{256x256}}} \\ \cline{2-7}
			&\quad \quad \quad {\textbf{ms}} &\quad \quad \quad \textbf{fps}  &\quad \quad \quad \textbf{ms} &\quad \quad \quad \textbf{fps} &\quad \quad \quad \textbf{ms} &\quad \quad \quad \textbf{fps} \\ \hline \hline  
			GAN-FCN~\citep{bi2018improving} &\quad \quad \quad 9 &\quad \quad \quad 120.64 &\quad \quad \quad 14 &\quad \quad \quad 87.62 &\quad \quad \quad 21 &\quad \quad 57.15 \\ 
			\textbf{Proposed MobileGAN} &\quad \quad \quad \textbf{5} &\quad \quad \quad \textbf{168.71} &\quad \quad \quad \textbf{8} &\quad \quad \quad \textbf{110.3} &\quad \quad \quad \textbf{14} &\quad \quad  \textbf{78.63} \\ \hline \hline 
		\end{tabular}
	}
\end{table}

In Table~\ref{table4_inference}, the inference time of MobileGAN was compared to GAN-FCN~\citep{bi2018improving}, which is a lightweight architecture, at different scales on input images ($64 \times 64$, $128 \times 128$ and $256 \times 256$). With $128 \times 128$ image scale, the inference time of  MobileGAN achieves is  8 ms (110 FPS) on a single GTX1080Ti GPU, while  GAN-FCN takes 6 ms  more than MobileGAN with the same configurations. As we can see, MobileGAN achieves  FPS higher than GAN-FCN at all scales of input images. In summary, MobileGAN is faster than the compared skin lesion segmentation models while having significantly better accuracy. Based on the computed inference time, it is obvious that the proposed MobileGAN model canbe run on a single Mobile GPU producing real-time and accurate skin lesion segmentation. 

\textbf{Results from the different variations of proposed model:} We assessed the effect of the addition PAM and CAM blocks on the baseline GAN model with and without the multiscale block.  Firstly, we assess the baseline GAN segmentation model (BL GAN). The $G$ network of BL GAN includes sequentially stacked factorized kernels in all convolution and deconvolution layers. As shown in Table~\ref{table5_abstudy1}, BL GAN obtains DSC and JSC scores of $83.61\%$ and $72.93\%$, respectively. 
Secondly, we add the PAM block to the BL GAN model (named as BL+PAM). Specifically, in the BL+PAM model, we added a PAM module after each downsampling and upsampling layers of both the encoder and decoder parts. We have also added a PAM module after the first downsampling layer in the discriminator network. The BL+PAM model obtains  DSC and JSC scores of $86.01\%$ and $75.96\%$, respectively.
Thirdly, we add the CAM block to the BL GAN model (BL+CAM). In the BL+CAM model, we added a CAM module after each factorized layer in the $G$ network. We also added a CAM module after the second downsampling layer in the discriminator network. The BL+CAM model gives DSC and JSC scores of $87.23\%$ and $76.65\%$, respectively. As we can see, the scores of BL+CAM are better than BL+PAM. Indeed, the addition of the CAM mechanism in the BL GAN model provides an efficient feature discriminability between skin lesion regions and normal skin region in the skin image.

\begin{table}[!t]
	\centering
	\caption{Different configurations of MobileGAN with the ISBI 2017 test dataset (bold refers to the top values).}
	\label{table5_abstudy1}
	\scalebox{0.84}{
		\begin{tabular}{@{}llllll@{}}
			\hline \hline
			\textbf{Methods} & \textbf{ACC} & \textbf{DSC} & \textbf{JSC} & \textbf{SEN} & \textbf{SPE} \\ \hline \hline 
			BL GAN & 95.63 & 83.61 & 72.93 & 79.42 & 96.91 \\ 
			BL+PAM & 96.72 & 86.01 & 75.96 & 82.48 & 98.40 \\ 
			\begin{tabular}[c]{@{}c@{}} {BL+CAM}\end{tabular} & {96.90} & {87.23} & {76.65} & {83.31} & {99.10} \\ 
			BL+PAM+CAM {w/o MS} & {97.25} & {88.87} & {78.76} & {85.49} & {99.59} \\ 
			MobileGAN &\textbf{97.61} & \textbf{90.63} & \textbf{81.98} & \textbf{87.81} & \textbf{99.92} \\ \hline \hline 
		\end{tabular}
	}
\end{table}
Furthermore, we add the PAM and CAM blocks to the BL GAN model without multiscale (BL+CAM+PAM w/o MS). This model obtains DSC and JSC scores of  $88.87\%$ and $78.76\%$, respectively.  As shown in the last row of Table~\ref{table5_abstudy1}, the addition of the multiscale, PAM, and CAM blocks to the BL GAN improves the DSC and JSC scores to $90.63\%$ and $81.98\%$, respectively. This last variation is the proposed MobileGAN model. Table~\ref{table5_abstudy1} demonstrates  MobileGAN outperforms all BL GAN variations in terms of the five metrics. Note that the all tested models are trained and tested with an input image of $128\times128$.

In general, the most of the related deep learning-based segmentation models downsample the input images to avoid the high computation of the deep models. For example, FrCN used an input image of $192\times256$, and SLSDeep used an image of $384\times384$. We demonstrate the effect of  different  resolutions of the input images ($64\times64$, $128\times128$, and $256\times256$) on the performance of the proposed MobileGAN model in Table~\ref{table6_abstudy2}. With the image size $64\times64$, the last layer of MobileGAN generates an $8\times8$ feature maps that yields very coarse level information, in which most of the important details are lost leading to low segmentation accuracy. With image size $256\times256$, the last layer of MobileGAN generates a feature map of $32\times32$, which also extracts skin lesions-irrelevant features (i.e., artifacts) that reduces the overall accuracy. In turn, the image size of $128\times128$ generates a $16\times16$ feature map at the last layer that retains skin lesions-relevant features and discards the irrelevant ones to achieve the best accuracy in terms of all metrics. Since our main goal is to achieve a lightweight model with keeping accurate skin segmentation results, we used the image size of $128\times128$ to train MobileGAN. 

\begin{table}[!t]
	\centering
	\caption{ The performance of MobileGAN with different input image resolutions of the ISBI 2017 test dataset.}
	\label{table6_abstudy2}
	\scalebox{0.84}{
		\begin{tabular}{@{}lllllllllll@{}}
			\hline \hline
			\textbf{Input size} & \multicolumn{2}{l} {\textbf{ACC}} & \multicolumn{2}{l} {\textbf{DSC}} & \multicolumn{2}{l}  {\textbf{JSC}} & \multicolumn{2}{l}  {\textbf{SEN}} & \multicolumn{2}{l}  {\textbf{SPE}}  \\ \hline \hline 
			64x64 & \multicolumn{2}{l} {94.44} & \multicolumn{2}{l} {87.59} & \multicolumn{2}{l} {77.76} & \multicolumn{2}{l} {84.16} & \multicolumn{2}{l} {95.36} \\ 
			128x128 &\multicolumn{2}{l} {\textbf{97.61}} & \multicolumn{2}{l} {\textbf{90.63}}& \multicolumn{2}{l} {\textbf{81.98}}& \multicolumn{2}{l} {\textbf{87.81}} & \multicolumn{2}{l} {\textbf{99.92}} \\ 
			256x256 & \multicolumn{2}{l}{96.72} & \multicolumn{2}{l}{89.72} & \multicolumn{2}{l}{79.49} & \multicolumn{2}{l}{86.36} & \multicolumn{2}{l}{98.21} \\ \hline \hline 
		\end{tabular}
	}
\end{table} 

\begin{table}[!b]
	\centering
	\caption{{Effect of different loss functions of on the performance of MobileGAN with the ISBI 2017 test dataset.}}
	\label{table7_abstudy3}
	\scalebox{0.84}{
		\begin{tabular}{@{}llllll@{}}
			\hline \hline
			\quad \quad \textbf{Loss} & \textbf{ACC} & \textbf{DSC} & \textbf{JSC} & \textbf{SEN} & \textbf{SPE} \\ \hline \hline 
			{MobileGAN+BCE} & {95.32} & {85.11} & {74.48} & {83.71}  & {98.02} \\ 
			MobileGAN+BCE+L1 & 96.90 & 87.26 & 76.80 & 85.05 & 99.30 \\
			{MobileGAN+BCE+Jaccard Loss} &  {96.97} & {89.56} & {79.88} & {86.90} & {99.49} \\
			MobileGAN+BCE+L1+ Jaccard Loss & \textbf{97.61} & \textbf{90.63} & \textbf{81.98} & \textbf{87.81} & \textbf{99.92} \\ \hline \hline 
		\end{tabular}
	}
\end{table}

\begin{figure}[!t]
	\centering
	\includegraphics[scale=0.6]{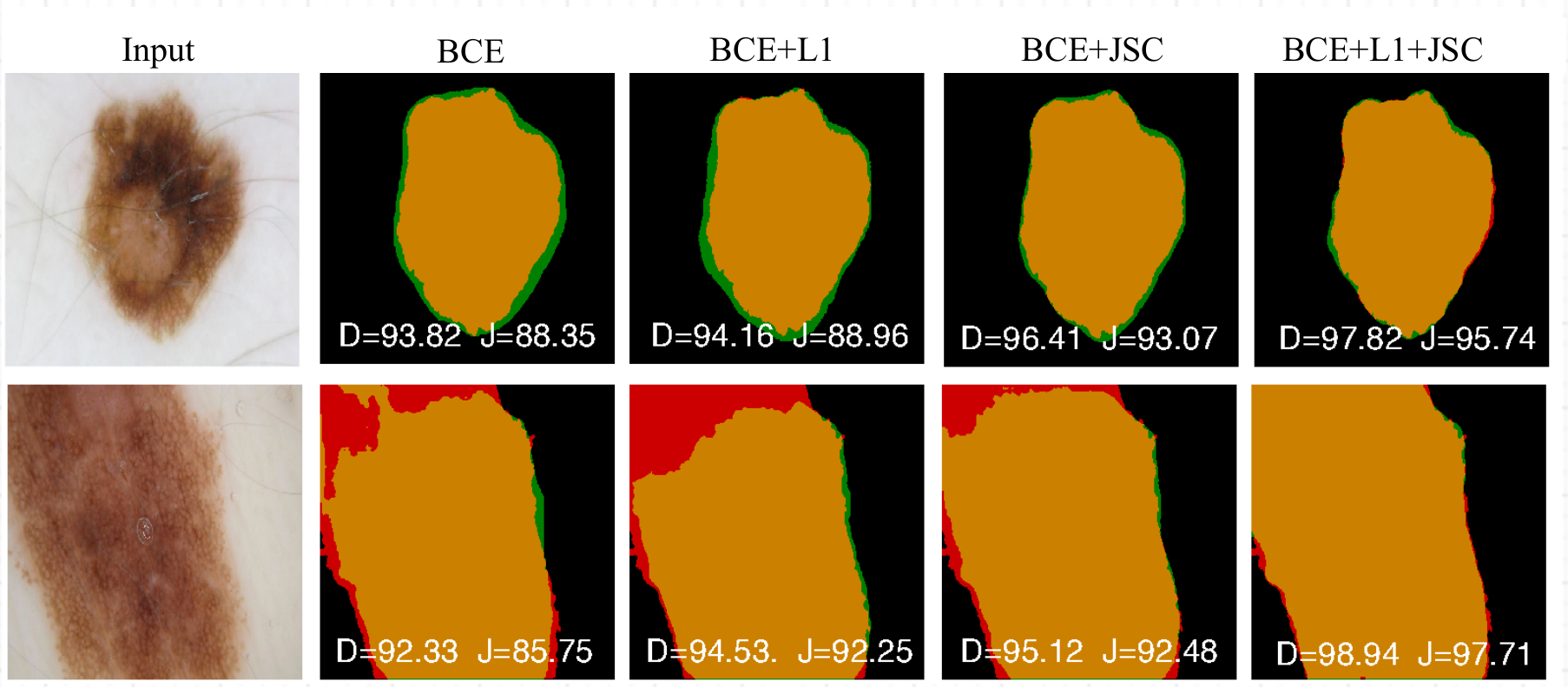}
	\caption{{Effects of different loss functions with MobileGAN on the test set of ISBI2017. Note that D and J refer to DSC and JSC, respectively.}}
	\label{fig22:lossfunction}
\end{figure}

Table~\ref{table7_abstudy3} shows the effect  of different loss function variations on the performance of  MobileGAN.  MobileGAN+BCE, MobileGAN+BCE+L1, MobileGAN+BCE+Jaccard and MobileGAN+BCE+L1+Jaccard loss obtain incremental JSC scores of $74.48\%$, $76.80\%$, $79.88\%$ and $81.98\%$, respectively. The proposed loss function shows significant improvement in terms of the five evaluation metrics. The use of the L1-loss function yields a reduction in the sensitivity to the choice of outliers. In turn, the use of Jaccard loss permits the MobileGAN to detect subtle abnormalities that cross-entropy loss did not detect. The combination of BCE+L1+Jaccard Loss with the proposed MobileGAN decreases the number of false positives from the segmented mask remarkably.  In Figure~\ref{fig22:lossfunction}, we present a couple of difficult samples from the ISBI 2017 dataset. We present the DSC and JSC scores on each mask, noting that D and J refer to DSC and JSC, respectively. The BCE+L1+Jaccard loss provides a JSC score of 95.74\% and 97.71\% with the upper and lower examples of Figure~\ref{fig22:lossfunction}, respectively. This analysis reveals the proposed loss function (BCE+L1+Jaccard loss) yields a better improvement at pixel-level in the segmentation result.

\section{Conclusions}
A lightweight and efficient GAN model for skin lesion segmentation, called MobileGAN, has been proposed in this paper. MobileGAN has been built by adapting a GAN model that comprises 1-D kernel factorized networks, multiscale aggregation, position, and channel attention mechanisms. MobileGAN has been assessed on the ISBI 2017 test and ISIC 2018 validation datasets. On the ISBI 2017 test dataset, it yields precise segmentation results with accuracy, sensitivity, specificity, Dice coefficient and Jaccard index, of $97.61\%$, $87.81\%$, $99.92\%$, $90.63\%$, and $81.98\%$, respectively. MobileGAN achieves a threshold JSC score of $78.4\%$ with the ISIC 2018 validation dataset. Comparing to the state-of-the-art, the number of parameters of MobileGAN has been significantly reduced, with only 2.35 million parameters.  In future work, we will implement a mobile application based on the MobileGAN model to segment skin lesions in images captured by low-resolution cameras.


\section*{Declaration of Competing Interest}
The authors declare that they have no conflict of interest.

\section*{Acknowledgment}
This research is funded by the program ``Marti Franque'' under the agreement between Universitat Rovira Virgili and Fundacio Catalunya La Pedrera, projects RTI2018-095232-B-C2, SGR 1742, CERCA, Nestore Horizon2020 SC1-PM-15-2017 (n 769643), EIT Health Validithi project. The authors acknowledge the support of NVIDIA Corporation with the donation of several Titan Xp GPU used for this research. M. Abdel-Nasser and D. Puig are partially supported by the Spanish Government under project DPI2016-77415-R.





\bibliographystyle{model2-names}\biboptions{authoryear}
\bibliography{mybibfile}

\end{document}